\documentclass[manuscript,nonacm]{acmart}

\AtBeginDocument{%
  \providecommand\BibTeX{{%
    \normalfont B\kern-0.5em{\scshape i\kern-0.25em b}\kern-0.8em\TeX}}}

\usepackage{multirow}
\usepackage{array}

\newcommand{\ainote}[1]{\textcolor{red}{[AI]:#1}}

\newcommand{\hnote}[1]{\textcolor{blue}{[H]:#1}}
\newcommand{\rnote}[1]{\textcolor{magenta}{[RK]:#1}}

\graphicspath{{./images/}} 

\usepackage{subcaption}

\usepackage{tikz}
\usepackage{calc}
\def\checkmark{\tikz\fill[scale=0.3](0,.35) -- (.25,0) -- (1,.7) -- (.25,.15) -- cycle;} 

\makeatletter
\let\@authorsaddresses\@empty
\makeatother

\begin{document}

\title{Leveraging GPT for the Generation of Multi-Platform Social Media Datasets for Research}


 \author{Henry Tari\textsuperscript{1}}
 \author{Danial Khan\textsuperscript{1}}
 \author{Justus Rutten\textsuperscript{1}}
 \author{Darian Othman\textsuperscript{1}}
 \author{Rishabh Kaushal\textsuperscript{1,2,3}}
 \author{Thales Bertaglia\textsuperscript{4}}
 \author{Adriana Iamnitchi\textsuperscript{1,2}}

 
 \affiliation{
 \institution{\\\textsuperscript{1}
 Department of Advanced Computing Sciences, Maastricht University}
 \city{Maastricht}
 \country{Netherlands}
 }

 \affiliation{
   \institution{\\\textsuperscript{2} Institute of Data Science,  Maastricht University}
   \city{Maastricht}
  \country{Netherlands}
 }
 
 \affiliation{
    \institution{\\\textsuperscript{3} Department of Information Technology, Indira Gandhi Delhi Technical University for Women}
   \city{Delhi}
  \country{India}
  }
 
 \affiliation{\\\textsuperscript{4}
 \institution{Utrecht University}
  \city{Utrecht}
  \country{Netherlands}
 }

\renewcommand{\shortauthors}{Tari et al.}

\begin{abstract}

Social media datasets are essential for research on disinformation, influence operations, social sensing, hate speech detection, cyberbullying, and other significant topics. However, access to these datasets is often restricted due to costs and platform regulations. As such, acquiring datasets that span multiple platforms—which are crucial for a comprehensive understanding of the digital ecosystem—is particularly challenging.
This paper explores the potential of large language models to create lexically and semantically relevant social media datasets across multiple platforms, aiming to match the quality of real datasets. We employ ChatGPT to generate synthetic data from two real datasets, each consisting of posts from three different social media platforms. We assess the lexical and semantic properties of the synthetic data and compare them with those of the real data.
Our empirical findings suggest that using large language models to generate synthetic multi-platform social media data is promising. However, further enhancements are necessary to improve the fidelity of the outputs.

\end{abstract}

\maketitle

\section{Introduction}

Social media platforms enabled the development of many socio-economical processes over the last decade. From connecting people worldwide to promoting goods and services through influencer marketing, these processes were analysed in quantitative and qualitative studies by a multi-disciplinary community that includes sociologists, political scientists, communication experts, economists, and computer scientists, among others. Such studies enabled progress in understanding individual and at scale behaviors, detecting malicious actors and operations, identifying political manipulation and disinformation campaigns, modeling information diffusion processes, etc. At the core of all these studies is data from social media platforms, which typically record how users interact with content and with each other. Never perfect~\cite{Tufekci_2014} and often costly to collect and curate~\cite{doi:10.1126/science.aaz8170}, social media datasets in general cannot be shared with the research community due to privacy and legal concerns even when social media posts are public. 

With the emergence of different social platforms with different types of content and audiences, it became clear that not any one platform is an accurate reflection of society on a particular topic at any time. Consequently, research shifted towards multi-platform studies of election discussions~\cite{10.1145/3328529.3328562,10.1145/3583780.3615121}, coordinated information operations~\cite{doi:10.1080/10584609.2019.1661889}, or cyberbullying~\cite{10.1007/s10579-020-09488-3}. However, datasets from multiple social media platforms are even more challenging to collect and curate, and much more difficult to re-collect for reproducibility studies or ``hydrate'' from message identifiers, as the individual platform challenges are compounded. For example, publicly available Twitter datasets include only the tweet IDs (and no actual content) are much more expensive to hydrate now due to policy changes at X related to its research API. Moreover, in most platforms, users remove their posts or accounts, or accounts are blocked: even when data collection is still practical, it is not guaranteed that the same data can be retrieved at a different time even when using the same collection parameters. 

This paper poses the following question: \emph{Given a dataset comprising text messages from various social media platforms on a specific topic within a certain time frame, how accurately can we generate a corresponding synthetic dataset?} Specifically, we examine a multi-platform social media dataset that includes messages posted publicly by a select group of users or pertaining to certain topics over a specified period. Our long-term goal is to facilitate reproducibility in social media research, as outlined in~\cite{schoch2023computational}, without contravening legal or platform-specific regulations. Ideally, researchers should be able to create and publicly share an accurate synthetic dataset derived from a multi-platform social media dataset gathered for their studies. This dataset could then be used with the same reliability as real data for various purposes, including training machine learning algorithms, testing tools on diverse datasets, and conducting reproducibility assessments. We hypothesize that this method will, in the long run, enhance transparency regarding content management on social media platforms, an initiative now supported by the EU Digital Services Act, thereby offering better protection against systemic risks. 

We take the first steps towards this vision by making the following contributions. 
First, we propose and evaluate two prompt engineering strategies appropriate for our objectives. Based on lexical and semantic comparisons between real and generated data, we propose a set of improvements for future work that can address observed limitations.  
Second, we quantify the fidelity of the synthetic datasets for six different social media platforms: Twitter, Instagram, Facebook, TikTok, YouTube and Reddit. 
We discovered that while convincing at individual social media post level in terms of appropriate use of lexical features such as hashtags and emojis, at collection level the (re)use of tags is unrealistically low.
Third, we show on two different datasets---one related to the US 2022 midterm elections and one on social media influencers---that chatGPT-generated datasets semantically reproduce the topics in the real data.




\section{Related Work}
Data driven machine learning requires large labeled training data.
However, obtaining high-quality labelled data is often a challenge.
Traditionally, numerous techniques \cite{li2022data} have been proposed to augment textual data, for instance, back-translation \cite{sennrich2015improving} and EDA \cite{wei2019eda}.
In this section, we summarize a recent trend whereby researchers have explored a variety of prompt design techniques on different LLMs to generate synthetic datasets to augment real datasets and improve downstream task performance. 

We begin with a focus on prior works that use social media data.
Moller et al.~\cite{moller2023prompt} found that synthetic data generated from GPT-4 and Llama-2 improves rare class performance in multi-class classification tasks. They evaluated their approach on ten tasks in computational social science. For generating synthetic data, they prompted LLMs by giving examples from single social media platform datasets.
Bertaglia et al.~\cite{bertaglia2024instasynth} generated synthetic Instagram posts using four prompting strategies to augment training data for the sponsored content detection task. They observed that the fidelity of the generated synthetic dataset and its utility for the downstream task can be conflicting. 
Ghanadian et al.~\cite{ghanadian2024socially} investigated the feasibility of generating synthetic Reddit posts for suicidal ideation. LLMs were prompted to generate suicidal text related to different factors like depression, anxiety, anger, hopelessness, and so on. They used two prompt approaches, namely, zero-shot (with no examples) and few-shot (with two examples for each factor).
They found that a mix of synthetic with real datasets gives the best results. 
Hartvigsen et al.~\cite{hartvigsen2022toxigen} created ToxiGen, a dataset of toxic and benign statements synthetically generated using GPT-3 by giving examples taken from Reddit in their prompt. They observed improvement in toxicity detection using ToxiGen data.
Similarly, Das et al.~\cite{das2024offlandat} proposed a dataset comprising implicit offensive speech targeted against 38 groups. They used zero-shot prompting grounded with positive intention to enable chatGPT to produce offensive text. Both ~\cite{hartvigsen2022toxigen} and ~\cite{das2024offlandat} employed human evaluation to assess the quality of the generated text.
Veselovsky et al.~\cite{veselovsky2023generating} focused on the sarcasm detection task using self-disclosed sarcastic tweets. They proposed three prompting approaches to generate faithful synthetic sentences, namely, grounding (by giving examples), taxonomy-based (theorizing a concept), and filtering (fine-tuning a model to distinguish synthetic from real sentences, and then using it on LLM outputs to filter out sentences labelled as synthetic). They found that grounding is most effective in detecting sarcasm.

Next, we summarize some recent works with a focus on learning different prompting techniques for synthetic data generation. 
Ubani et al.~\cite{ubani2023zeroshotdataaug} used chatGPT with zero-shot prompting to create synthetic data for low-resource settings. They evaluated on three datasets related to movie reviews, spoken natural language for personal services, and question classification. 
Dai et al.~\cite{dai2023auggpt} proposed an approach for text augmentation by using single-turn and multi-turn dialogues in chatGPT. They rephrased the input sentence into multiple sentences which are similar in concept but different in semantics. They validated their approach on three datasets - Amazon customer reviews, common medical symptoms descriptions, and scientific abstracts.
Aboutalebi et al.~\cite{aboutalebi2024magid} introduced an approach to generate synthetic multi-modal datasets. They provide a text-only dialogue (sequence of utterances) to LLMs, and prompt LLMs to identify the most suitable textual utterance that can be augmented with an image so that the dialogue is more engaging. They used zero-shot, few-shot, and chain of reasoning techniques to identify the most suitable utterance. 
Josifoski et al.~\cite{josifoski2023exploiting} used LLMs with zero-shot and few-shot prompting to generate input sentences for a known target triplet comprising of subject, object, and their relation from well known Wikidata knowledge graph dataset.

These prior works focus mainly on the utility of the resulting synthetic dataset for solving a downstream task.
As a consequence, their prompting approach is often biased towards a specific downstream task.
In addition, some prior works do use multiple datasets to valid their approach, but these datasets are often unrelated to social media platforms. 
To the best of our knowledge, no prior work has worked on multi-platform social media data.
In contrast, our approach in this paper is to focus on generation of high-fidelity synthetic datasets for multi-platform social media data without biasing it a particular downstream task. 

\section{Datasets}


For this study, we selected two multi-platform datasets, each containing messages posted by users on three different platforms. 
The first is a subset of the dataset collected by Aiyappa et al.~\cite{Aiyappa2023} on the topic of the US elections 2022 from three platforms: Twitter, Facebook and Reddit. 
The second consists of posts on TikTok, Instagram and YouTube by a group of social media influencers registered with the Dutch government on advertising, sponsorship, and product placement.\footnote{Content creators (influencers) with more than 500k followers are required to register with the Dutch Media Authority as of July 1st, 2022~\cite{MinOCW2022}.} 

These two datasets cover a set of characteristics that enable a wider exploration in the space of synthetic social media datasets. 
First, both are multi-platform datasets, connected by either users authoring posts on multiple platforms (the Dutch influencers) or by topic (the US 2022 elections). 
Second, they together cover six popular social media platforms. 
Third, their topics and style of discussions should be quite different in terms of diversity measures. 
On the one hand, the US election dataset should be more homogeneous in topic yet more heterogeneous in the authorship type since it theoretically contains posts from users with various lexical styles and political leanings. 
On the other hand, the Dutch influencer dataset should be more diverse in topics (e.g., due to the different brands and images promoted), and potentially more homogeneous in authorship style, since this user group has the common goal of monetizing their content. 
Indeed, here are two captions from different users on TikTok. 
Although the topics are different, there is a similarity in the writing style: \textit{``EASY CARDIGAN HACK SAVE FOR LATER GIRLS \#cardigan \#cardiganoutfit \#tutorial \#howtostyle \#fashionhacks''} \& \textit{``PlayStation Lifehack on Xbox! \#PS5 \#Xbox \#Gaming \#Gamer \#Yorrick''}

Because our focus is on textual data only (thus, ignoring user identities, user interactions, and the timing of the posts that are important in social media analysis for different objectives) and on evaluating the efficacy of LLM's generative capability with few shot grounding samples, we randomly selected 1000 posts from each platform. Given their different focus on content modalities, posts from different platforms are highly distinct in length and usage of textual features such as hashtags and URLs. 
Hashtags are hardly used in Reddit but common in Instagram and TikTok; user tags are highly popular on YouTube and practically non-existent on Reddit; and Reddit messages---often spanning multiple paragraphs--- are much longer than TikTok or Instagram posts, which consist of captions to images or short movies. 

\section{Synthetic Data Generation via chatGPT}
\label{sec:promt}

To generate data, we opted for OpenAI's model GPT-3.5-turbo, which is a text completion model capable of retaining context from previous interactions. It is also the most cost-effective\footnote{https://openai.com/pricing} option among OpenAI's model offerings, making it preferable for generating larger datasets. Another factor that influenced our choice of this model is its compact size compared to other options, which helps reduce the carbon footprint of our approach, especially since we are generating large multi-platform datasets. 

\begin{figure}[ht]
\centering
{\includegraphics[width=0.8\textwidth]{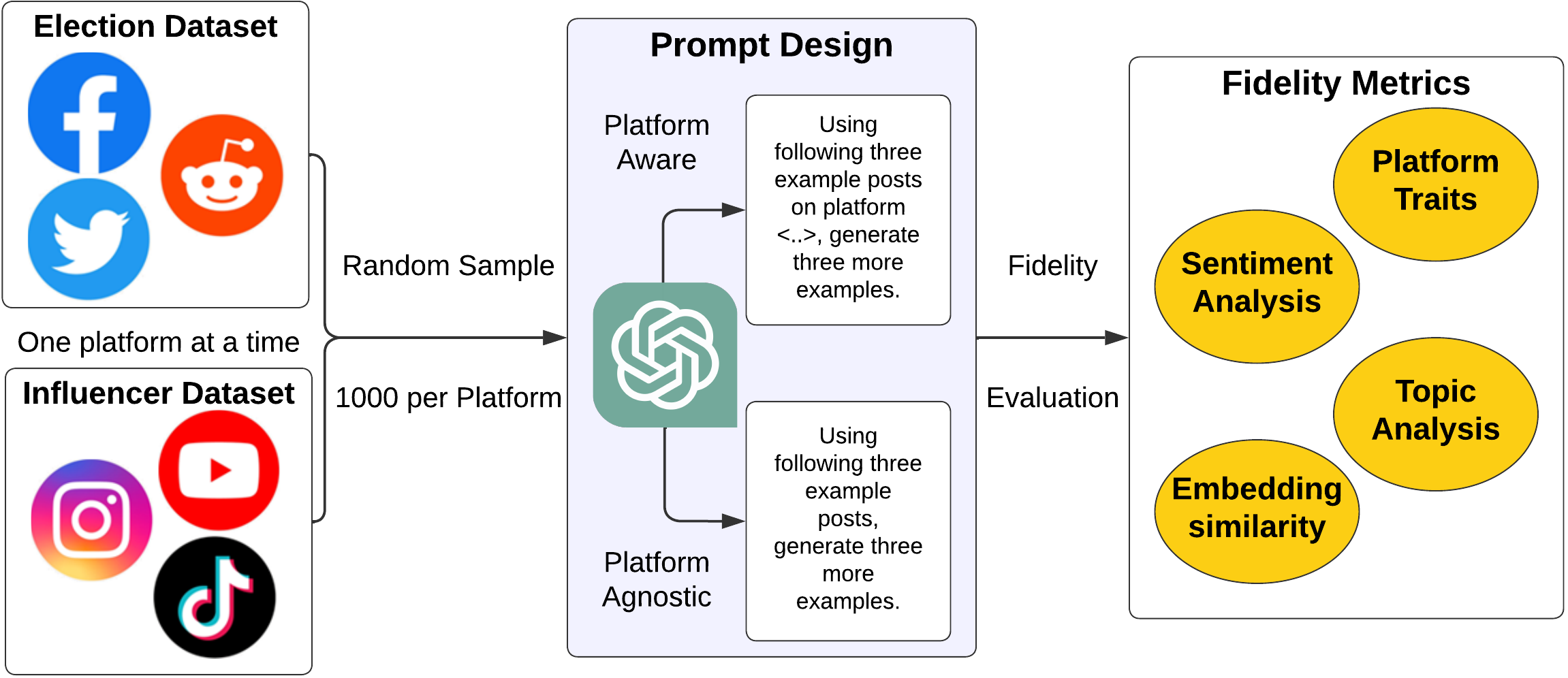}}%
\caption{Pipeline of our methodology for generating and evaluating synthetic social media datasets.}
\label{fig:prompt_struct}
\end{figure}
\noindent 
We employed various strategies to generate text resembling realistic social media posts.
We began with a basic zero-shot prompt instructing the generation of a specific social media post such as, ``\textit{Generate a Twitter post based on US elections 2022}'', but it proved ineffective in producing diverse results: Upon receiving multiple iterations of the same prompt, chatGPT starts reproducing the same outputs with only minor variations such as adding an emoji or using a different hashtag at the end.
As a consequence, we tried few-shot prompting by providing a variety of contextual examples in prompts.
We experimented with prompts ranging from two to five input examples. 
Generally, using more examples enhanced the diversity and variation in the output, which seemed particularly effective for platforms with shorter messages, such as Twitter. After experimenting with different numbers of grounding examples, across all the platform, we obtained the best results when providing three randomly chosen posts from our real dataset and prompting chatGPT to generate three output posts. Using more than three examples becomes problematic when dealing with platforms like Reddit, where the long text messages are challenging for the token limit of a single API call and for keeping track of context. 

Figure~\ref{fig:prompt_struct} depicts the pipeline of our prompt approach.
We implemented our prompting approach using the OpenAI API\footnote{https://platform.openai.com/docs/introduction}. 
Each prompt specified a desired JSON output format to facilitate the extraction of the generated posts.
Outputs that deviated from the specified JSON format were discarded, as the model occasionally produced invalid results. 
During our initial prompt experience, we observed that chatGPT often ignored the context of the three input examples provided. 
Therefore, we refined the prompt instruction and included phrases like `Using following examples ...', `Based on the examples below ...', and so on. 
This modification improved performance within the given context, aligning results more closely with our real dataset. 
However, after some runs, the chatGPT model often produced results that were mere reproductions of the examples.
Henceforth, we further rephrased our prompt instruction by adding phrases like `generate new posts keeping them true to their content and writing style.'
This prompt instruction encouraged the generation of diverse and novel social media posts, resembling the examples but distinct enough to be considered unique. 
Throughout this process, we observed that the model tended to ignore instructions when the size of the examples was large, likely due to limited attention capacity. To address this issue, we placed a single-line summary of the instructions at the end of the prompt.


In the end we converged to two distinct types of prompting, referred to as \emph{`platform aware'} and \emph{`platform agnostic'}.
In \emph{platform-aware prompting}, we explicitly ask in the prompt instructions to generate text for the target social media platform. 
This prompting strategy was adopted in the hope of leveraging the model's capability to generate posts with styling characteristics that are typical of the target social media platform.
In \emph{platform-agnostic prompting}, the chatGPT model was not informed about the target social media platform, but it was asked to keep the generated output similar in content and length to the examples provided.

We experimented with different values for the model's two primary parameters, $temperature$ (T) and $top\_p$ (P). 
$T$ values exceeding 1 yielded highly varied yet extremely artificial results, while those below 0.7 mostly replicated the given examples. 
$T$ values at 0.7 and 1, however, produced realistic variations, which were then selected for further exploration. 
Likewise, the $P$ values exhibited a comparable trend. 
Consequently, values of 1 and 0.7 for both $T$ and $P$ were chosen for further experimentation. 
When both parameters were set to 0.7, sub-optimal results were obtained, and hence this combination was discarded from further experiments.
Eventually, we perform experiments on three combinations of $T$ and $P$ values, namely, <$T$=0.7, $P$=1>, <$T$=1, $P$=0.7>, and <$T$=1, $P$=1>.

\section{Evaluation of Synthetic Social Media Posts}
\label{sec:results}
We evaluate the textual fidelity of the synthetic posts compared to real posts.
We measure fidelity along four dimensions. 
First, we compare the occurrence of platform-specific textual features typical to social media posts, namely hashtags, user tags, URLs, and emojis.
Second, we compare the sentiment of synthetic to real posts. 
Third, we assess the diversity and coherence of topics found in synthetic posts compared to real posts.
And finally, we measure text similarity between real and synthetic posts in the embedding space.

\subsection{Hashtags, User Tags, URLs and Emojis in Synthetic Data}

Table~\ref{tab:occurentokens} presents the average number of hashtags, user tags, URLs, and emojis, respectively, per real and synthetic post, along with distinct occurrences of these tokens. 
On average, chatGPT tends to create more hashtags for Twitter, Facebook, YouTube and even Reddit, where the use of hashtags in reality is rather low. 
However, hashtags are underrepresented in the synthetic data for Instagram and TikTok, whereas in our Dutch influencers dataset hashtags are heavily used for monetization. 
In terms of hashtag diversity, we observe a larger number of distinct hashtags in the synthetic data than in the real (shown between parentheses in Table~\ref{tab:occurentokens}), indicating lower hashtag reuse among posts in the synthetic datasets than in reality.  
On social media platforms, hashtags are meant to connect posts with similar topics, yet,  understandably, chatGPT misses this point. 
This difference was also observed in~\cite{bertaglia2024instasynth} and likely relates to the sequential manner in which LLMs work, without trying to make connections between individual outputs. 
However, this may also suggest that prompting strategies that emphasize shared hashtags among the examples given in prompts could lead to more realistic generated datasets.

\begin{table}[!h]
\caption{The average number of lexical features per message in the real and synthetic social media datasets. The number of distinct tokens appears within parentheses. Each data collection includes 1000 social media posts from each platform.}
\resizebox{\columnwidth}{!}{%
\begin{tabular}{@{}llrrrrrrr@{}}

\toprule
\textbf{Platform} & \textbf{Token} & \textbf{Real} & \multicolumn{3}{c}{\textbf{Platform Agnostic}} & \multicolumn{3}{c}{\textbf{Platform Aware}} \\ \cmidrule(lr){4-6} \cmidrule(lr){7-9}
 &  &  & \textbf{T=0.7 P=1} & \textbf{T=1 P=0.7} & \textbf{T=1 P=1} & \textbf{T=0.7 P=1} & \textbf{T=1 P=0.7} & \textbf{T=1 P=1} \\ \midrule
\multirow{4}{*}{Twitter} & Hashtag & 1.19 (964) & 1.77 (1402) & 1.74 (1403) & 1.84 (1458) & 1.73 (1378) & 1.69 (1304) & 1.81 (1405) \\
 & Tag & 1.02 (930) & 0.72 (680) & 0.70 (654) & 0.56 (516) & 0.65 (608) & 0.60 (555) & 0.54 (510) \\
 & URL & 0.35 (356)& 0.13 (126) & 0.12 (123) & 0.07 (75) & 0.06 (62) & 0.07 (69) & 0.05 (50) \\
 & Emoji & 0.30(175) & 0.34 (183) & 0.32 (186) & 0.43 (225) & 0.31 (163) & 0.26 (143) & 0.37 (193) \\ \midrule
\multirow{4}{*}{Facebook} & Hashtag & 0.92 (830) & 1.74 (1420) & 1.78 (1420) & 1.75 (1381) & 1.28 (1080) & 1.35 (1114) & 1.38 (1118) \\
 & Tag & 0.05 (48) & 0.01 (16) & 0.19 (22) & 0.03 (36) & 0.01 (16) & 0.02 (19) & 0.01 (9) \\
 & URL & 0.17 (190) & 0.04 (44) & 0.05 (56) & 0.04 (38) & 0.04 (39) & 0.04 (44) & 0.03 (32)\\
 & Emoji & 0.00 (3) & 0.10 (78) & 0.08 (55) & 0.15 (99) & 0.11 (22) & 0.07 (58) & 0.17 (95) \\ \midrule
\multirow{4}{*}{Reddit} & Hashtag & 0.14 (72) & 1.24 (963) & 1.15 (830) & 1.15 (908) & 0.13 (118) & 0.10 (94) & 0.12 (128) \\
 & Tag & 0.00 (2) & 0.00 (2) & 0.00 (1) & 0.03 (4) & 0.00 (1) & 0.00 (1) & 0.00 (1) \\
 & URL & 0.09 (110) & 0.01 (11) & 0.01 (16) & 0.01 (14) & 0.01 (14) & 0.02 (19) & 0.01 (14) \\
 & Emoji & 0.01 (3) & 0.08 (45) & 0.04 (31) & 0.11 (63) & 0.02 (22) & 0.01 (14) & 0.02 (16) \\  \midrule
 \hline
\multirow{4}{*}{Instagram} & Hashtag & 3.64 (2058) & 2.34 (1835) & 2.39 (1874)& 2.20 (1763) & 2.65 (2115) & 2.69 (2133) & 2.51 (2122) \\
 & Tag & 0.77 (539) & 0.26 (250) & 0.27 (254) & 0.26 (245) & 0.28 (272) & 0.27 (260) & 0.29 (274) \\
 & URL & 0.01 (8) & 0.00 (2) & 0.00 (2) & 0.00 (2) & 0.00 (1) & 0.00 (1) & 0.00 (1) \\
 & Emoji & 1.40 (576) & 1.64 (558) & 1.62 (527) & 1.63 (553) & 1.90 (585) & 1.92 (556) & 1.98 (563) \\ \midrule
\multirow{4}{*}{TikTok} & Hashtag & 3.41 (1805) & 2.80 (1964) & 2.76 (1909) & 2.77 (2031) & 3.16 (2069) & 3.19 (2086) & 3.21 (2166) \\
 & Tag & 0.20 (173) & 0.10 (95) & 0.10 (98) & 0.10 (99) & 0.10 (97) & 0.11 (107) & 0.12 (113) \\
 & URL & 0.00 (0) & 0.00 (0) & 0.00 (0) & 0.00 (0) & 0.00 (0) & 0.00 (0) & 0.00 (0) \\
 & Emoji & 1.00 (402) & 1.56 (525) & 1.66 (539) & 1.72 (561) & 1.72 (545) & 1.70 (528) & 1.76 (553) \\ \midrule
\multirow{4}{*}{YouTube} & Hashtag & 1.02 (499) & 2.25 (1291) & 2.28 (1296) & 2.15 (1334) & 2.02 (1245) & 2.06 (1263) & 1.96 (1211) \\
 & Tag & 1.00 (82) & 0.05 (52) & 0.05 (46) & 0.05 (46) & 0.04 (41) & 0.05 (52) & 0.04 (41) \\
 & URL & 8.67 (2489) & 0.34 (229) & 0.37 (257) & 0.40 (288) & 0.98 (613) & 1.11 (726) & 0.96 (589)  \\
 & Emoji & 1.67(141) & 0.91 (297) & 0.78 (285) & 0.95 (334) & 1.56 (435) & 1.31 (420) & 1.55 (432) \\ \bottomrule
\end{tabular}%
}
\label{tab:occurentokens}
\end{table}
Tagging other platform users is common on Twitter, Instagram and YouTube and almost non-existent on Reddit. 
In our experiments, chatGPT underestimates the use of user tags in all cases, except for Reddit, where it correctly estimates virtually no use of such tags. 
The same pattern of no tag reuse is evident. 
For example, in the case of YouTube influencers, where on average every post out of the 1000 in our dataset tags one out of 82 distinct users, in synthetic posts each tagged user is distinct from the users tagged in other posts. 
This time, due to generating fewer user tags on average per post, chatGPT also generates fewer unique such tags than there are in our datasets. 

The number of URLs included in posts is also underrepresented in synthetic data, most notably for YouTube, where influencers typically include their accounts on other platforms via web addresses. 
However, due to the relatively low number of URLs present in both datasets (except the YouTube video descriptions), the difference between real data and synthetic data in this metric is limited. 

Finally, emojis are accurately represented in the synthetic datasets for most platforms overall.
Remarkably, there is a clear distinction between the platforms represented by our two datasets: the Dutch influencers use many emojis, while in the US election conversations there are significantly fewer. 
While unclear if the topics or the platform practices are determining this difference, it is notable that chatGPT accurately captures this difference. 
Again, synthetically generated posts include a much larger variety of emojis than in our real data, especially for the more restrained platforms, such as Facebook and Reddit. 

While we experimented with two prompting strategies and three combinations of parameters each, we could not see a consistent benefit of any one combination in these lexical metrics. 

\subsection{Sentiment Analysis} 
Moving beyond lexical tokens, we compare sentiments expressed in generated content with real data. 
We employed the \emph{cardiffnlp/twitter-roBERTA-base-sentiment-latest} model proposed by Loureiro et al.~\cite{loureiro-etal-2022-timelms} available at Hugging Face\footnote{https://huggingface.co/cardiffnlp/twitter-roberta-base-sentiment-latest} to classify posts as \textit{positive}, \textit{negative}, or \textit{neutral}.
The model is based on the RoBERTa architecture~\cite{Liu2019RoBERTaAR} and is pre-trained for language modelling on ~124M tweets and then fine-tuned for sentiment analysis on the TweetEval benchmark~\cite{barbieri2020tweeteval}.
Table~\ref{tab:sent_inf_ds} presents the percentage of posts belonging to a particular sentiment category for each platform.
We observe that generated content is more positive and less negative than reality for all platforms.
Neutral sentiment is also under-expressed in synthetic content across all platforms except for Reddit.
Recent work~\cite{das2024offlandat, dutta2023down, hartvigsen2022toxigen,zhang2023comprehensive} has shown the use of LLMs for generating toxic content. 
Therefore, our observed outputs from the chatGPT suggest that efforts are being made to decrease negativity in the generated content.
However, from the standpoint of improving fidelity, prompting techniques that use positive intent as presented in~\cite{das2024offlandat} could generate content that better matches the negative sentiment in real posts.

Again, the prompting strategy and the parameters used do not yield any consistent lessons. 
The relative tone of the two datasets is maintained for all settings: the synthetic US election posts are classified as more negative than the synthetic posts generated from the Dutch influencers examples, which accurately reflects the real datasets, independent of prompting and GPT parameters.
In the rest of our evaluation section we will only include the default parameters (temperature $T=1$, probabilistic sampling $P=1$) for the two prompting strategies.  

\begin{table}[H]
\caption{Percentage of posts out of a collection of 1000 for each platform and setting that are classified based on sentiment analysis.}
\begin{tabular}{@{}llrrrrrrr@{}}
\toprule
\textbf{Platform} & \textbf{Sentiment} & \textbf{Real} & \multicolumn{3}{c}{\textbf{Platform Agnostic}} & \multicolumn{3}{c}{\textbf{Platform Aware}} \\ \cmidrule(lr){4-6} \cmidrule(lr){7-9}
 &  &  & \textbf{T=0.7 P=1} & \textbf{T=1 P=0.7} & \textbf{T=1 P=1} & \textbf{T=0.7 P=1} & \textbf{T=1 P=0.7} & \textbf{T=1 P=1} \\ 
 \hline
Twitter & Negative & 46.76 & 32.87 & 33.30  & 26.75 & 31.86 & 31.85 & 29.50 \\              
    & Neutral  & 21.06  & 17.93  & 15.91  & 15.50     & 15.48   & 16.37   & 13.50  \\
    & Positive                      & 32.18                       & 49.20                                                                            & 50.79                                                                             & 57.75                                                                         & 52.66                                                                                 & 51.78                                                                                 & 57.00                                                                               \\ 
                      
                         \hline
                                      & Negative                      & 34.19                       & 22.74                                                                            & 24.44                                                                             & 19.77                                                                         & 21.98                                                                                 & 23.47                                                                                 & 18.97                                                                               \\
Facebook                              & Neutral                       & 31.80                       & 29.73                                                                            & 29.77                                                                             & 27.28                                                                         & 26.57                                                                                 & 26.95                                                                                 & 27.51                                                                               \\
                                      & Positive                      & 34.01                       & 47.53                                                                            & 45.79                                                                             & 52.95                                                                         & 51.45                                                                                 & 49.58                                                                                 & 53.52                                                                               \\ \hline 
                                      & Negative                      & 81.91                       & 57.54                                                                            & 56.54                                                                             & 56.03                                                                         & 65.10                                                                                 & 62.93                                                                                 & 60.95                                                                               \\
Reddit                                & Neutral                       & 7.95                        & 15.25                                                                            & 15.94                                                                             & 14.56                                                                         & 15.29                                                                                 & 17.45                                                                                 & 18.43                                                                               \\
                                      & Positive                      & 10.14                       & 27.21                                                                            & 27.52                                                                             & 29.41                                                                         & 19.61                                                                                 & 19.62                                                                                 & 20.62                                                                               \\ \hline \midrule
                                      & Negative                      & 3.3                       & 1.9                                                                            & 2.0                                                                             & 1.3                                                                         & 0.9                                                                                 & 0.6                                                                                 & 0.7                                                                               \\

Instagram                 & Neutral                    & 29.2                                        & 4.7                              & 5.4                              & 4.7                            & 4.3                              & 5.0                                & 4.6                            \\
                          & Positive                   & 67.5                                        & 93.4                             & 92.6                             & 94.0                           & 94.8                             & 94.4                             & 94.7                           \\ \hline
                          & Negative                   & 7.1                                         & 4.6                              & 4.7                              & 5.1                            & 6.2                              & 7.1                              & 5.8                            \\
TikTok                    & Neutral                    & 39.0                                        & 9.4                              & 9.5                              & 10.8                           & 14.4                             & 15.8                             & 14.7                           \\
                          & Positive                   & 53.9                                        & 86.0                             & 85.5                             & 84.1                           & 79.4                             & 77.1                             & 79.5                           \\ \hline
                          & Negative                   & 2.0                                         & 1.0                              & 0.1                              & 1.1                            & 1.0                              & 0.7                              & 0.7                            \\
YouTube                   & Neutral                    & 40.0                                        & 8.3                              & 8.7                              & 6.3                            & 8.0                              & 8.9                              & 7.0                            \\
                          & Positive                   & 58.0                                        & 90.7                             & 90.6                             & 92.6                           & 91.0                             & 90.4                             & 92.3                           \\ \hline
\end{tabular}
\label{tab:sent_inf_ds}
\end{table}

\subsection{Topic Generation and Overlap}
Considering that social media posting behavior revolves around topics, we analyze topics in the synthetic data and compare them with topics in real posts. 
Intuitively, the generated data should be representative of the midterm US 2022 elections and respectively the themes that influencers choose to talk about, yet new topics in these general subjects are welcome for diversity and potential utility of synthetic data.

\begin{figure}[!htb]
\centering
\begin{subfigure}{0.16\textwidth}
\includegraphics[width=1.3\linewidth,trim={5cm 1.8cm 1.3cm 0},clip]{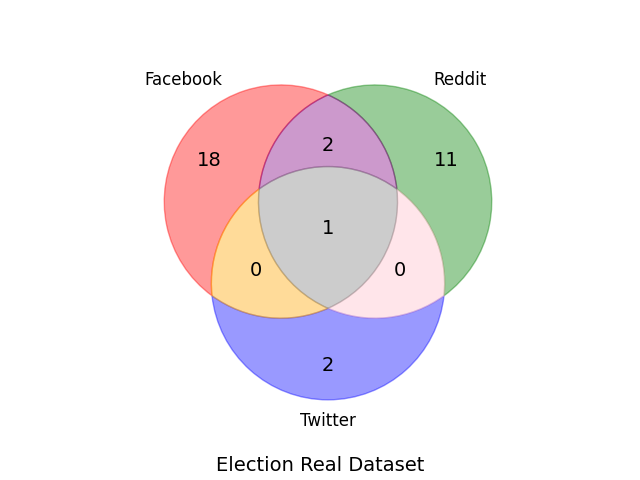}
    \caption{Real (US)}
\end{subfigure}
\begin{subfigure}{0.16\textwidth}
\includegraphics[width=1.3\linewidth,trim={5cm 1.8cm 1.3cm 0},clip]{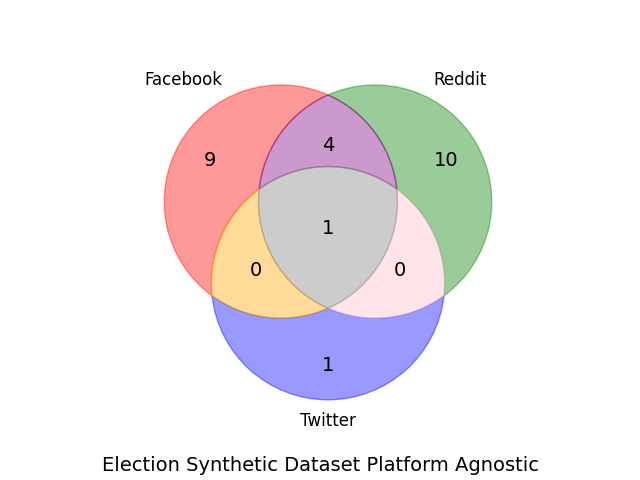}
    \caption{Agnostic (US)}
\end{subfigure}
\begin{subfigure}{0.16\textwidth}
\includegraphics[width=1.3\linewidth,trim={5cm 1.8cm 1.3cm 0},clip]{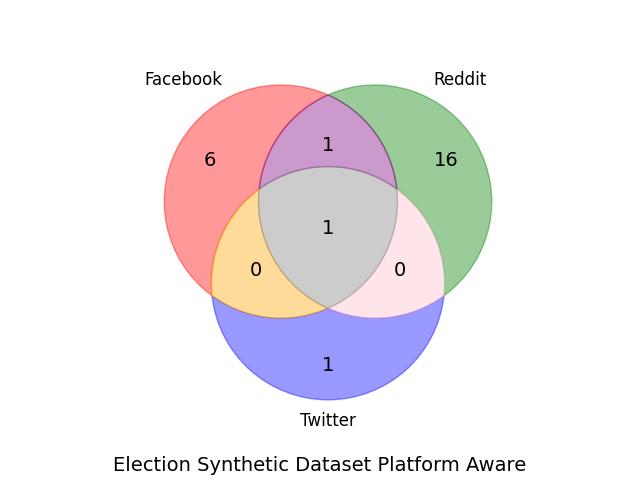}
    \caption{Aware (US)}
\end{subfigure}
\begin{subfigure}{0.16\textwidth}
\includegraphics[width=1.3\linewidth,trim={4.8cm 1.8cm 1.1cm 0},clip]{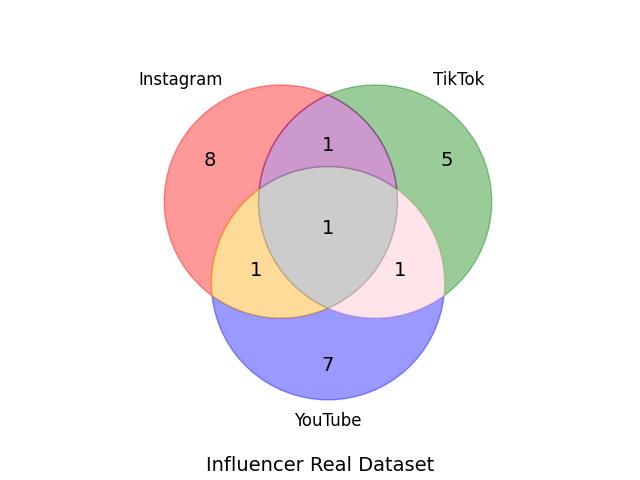}
    \caption{Real (NL)}
\end{subfigure}
\begin{subfigure}{0.16\textwidth}
\includegraphics[width=1.3\linewidth,trim={4.8cm 1.8cm 1.1cm 0},clip]{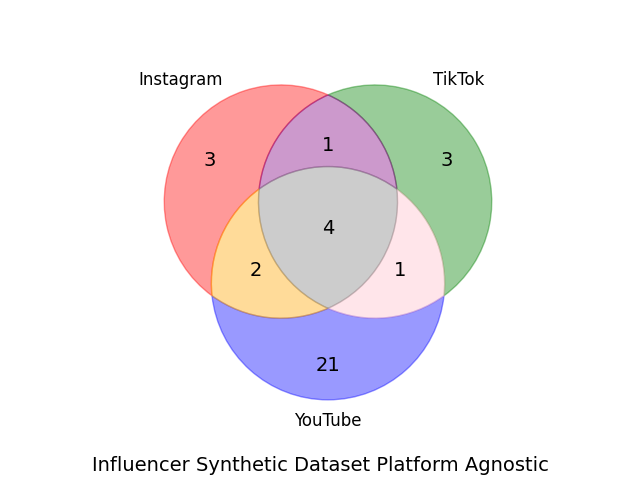}
    \caption{Agnostic (NL)}
\end{subfigure}
\begin{subfigure}{0.16\textwidth}
\includegraphics[width=1.3\linewidth,trim={4.8cm 1.8cm 1.1cm 0},clip]{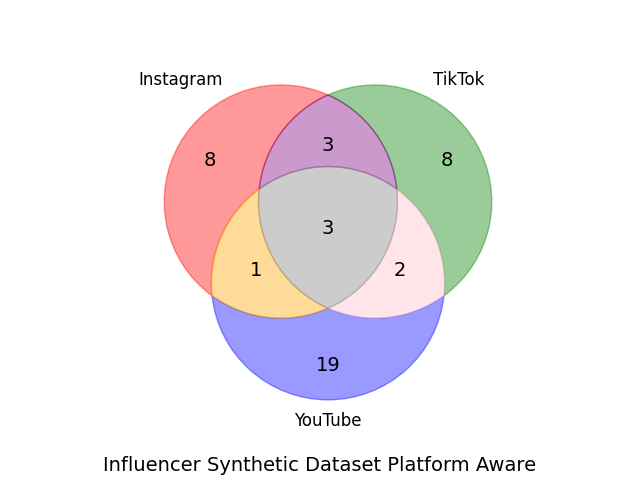}
    \caption{Aware (NL)}
\end{subfigure}
\caption{Topic overlap among platforms in the real and synthetic datasets. Platform-agnostic/aware prompts, $P=1$, $T=1$.}
\label{fig:inf_ele_real_syn}
\end{figure}

For topic extraction we used BERTopic~\cite{grootendorst2022bertopic} which is being increasingly used for social media data~\cite{verbeij2024happiness,hanley2023happenstance,hua2022using}.
For sentence embedding required by BERTopic, we used all-MiniLM-L6-v2\footnote{https://huggingface.co/sentence-transformers/all-MiniLM-L6-v2}~\cite{wang2020minilm} sentence transformer that works with sentences as well as short paragraphs, which is useful in our case because of the length variety from long Reddit and YouTube posts to short tweets.
We performed minimal text cleaning for optimization purposes. 
Salient features such as hashtags and emojis were abundant in the influencer dataset, and omitting them resulted in posts losing their meanings. 
Therefore, only URLs and stop words were removed during this process.
We focused only on the topics that appeared in at least 10 social media posts on the corresponding platform.
Figure~\ref{fig:inf_ele_real_syn} shows the number of shared and disjoint topics across different platforms in real and synthetic datasets. 
For topic overlap, we applied Open AI's embedding model~\cite{neelakantan2022text} (text-embedding-3-large) on top-10 most representative words for each topic and used cosine similarity for topic comparison with a threshold obtained empirically.


In the elections dataset of real posts, there is more topic diversity on Facebook and Reddit than on Twitter. 
Both prompting approaches can replicate the topic diversity, though not as much in the case of Facebook. Platform-aware prompts generate more topics in Reddit, perhaps because Reddit post lengths are the highest among all platforms, giving chatGPT more grounding. 
The topic `Voter Engagement' appears common in all platforms, both in real and synthetic (platform agnostic) data.
`Women Reproduction Rights' appears as a common topic across Facebook and Reddit. 
Platform agnostic prompting generates four topics of which two (`Mail-in Voting' and `Women Reproduction Rights') are the same as those present in the real posts.
In the influencer dataset of real posts, we observe that topics related to health, wellness, and fitness occur across all platforms. 

\begin{figure}[htb]
    \centering
    \includegraphics[width=0.75\linewidth]{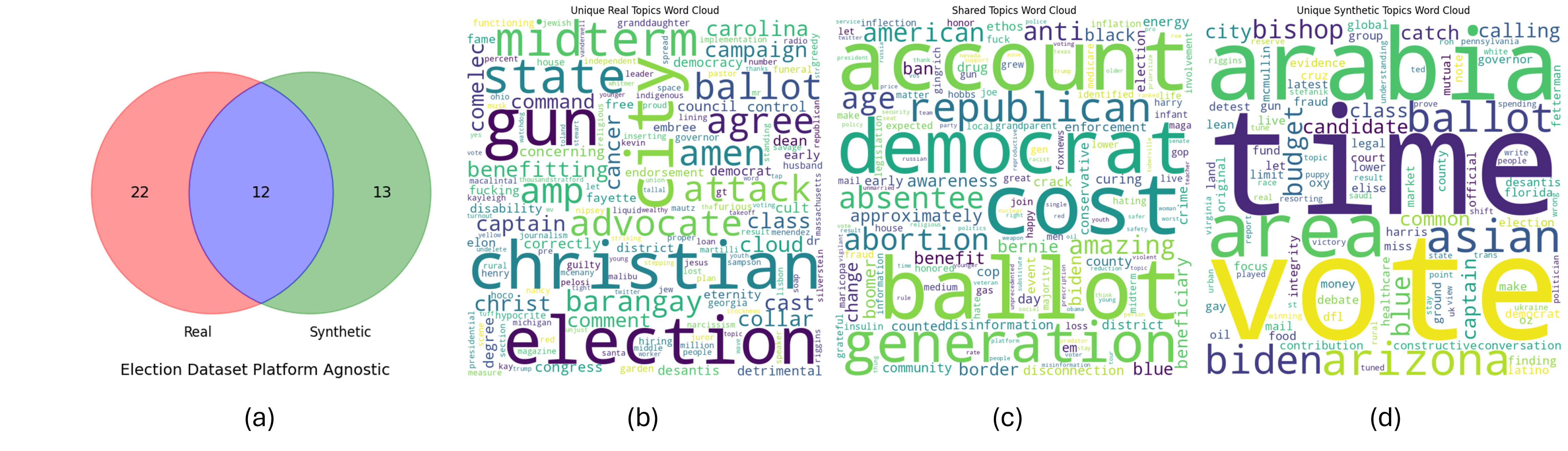}
    \caption{Topic overlap between real and synthetic (platform agnostic) data on the US elections~(a). Word clouds of unique topics in the real dataset~(b), common topics~(c), and new topics in the synthetic data~(d) on US elections.}
    \label{fig:EDC_plt_ag}
\end{figure}

\begin{figure}[htb]
    \centering
    \includegraphics[width=0.75\linewidth]{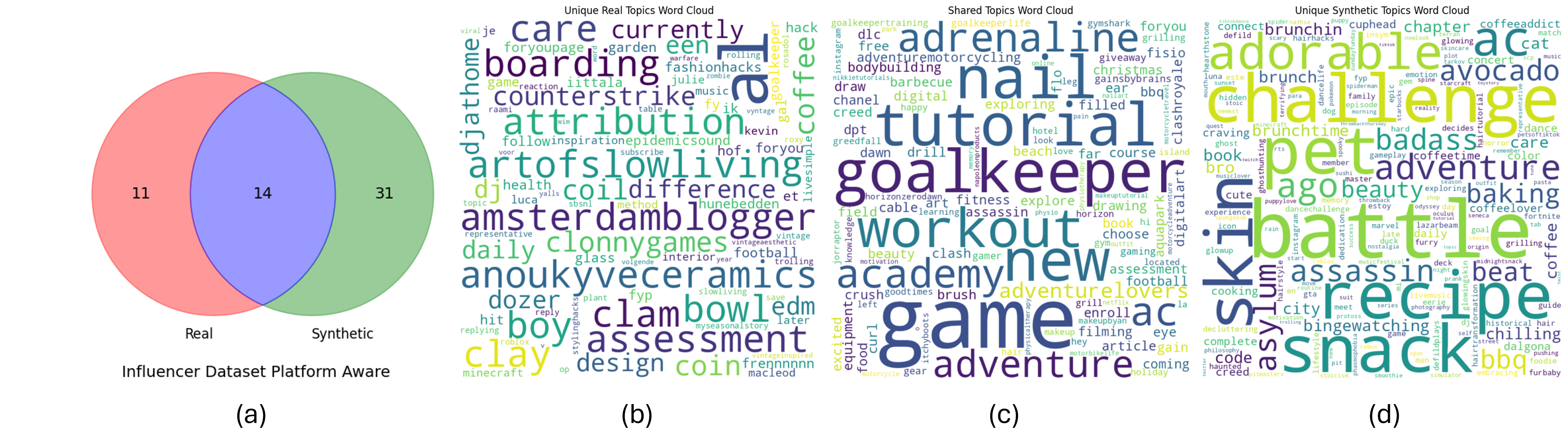}
    \caption{Topic overlap between real and synthetic (platform aware) data on the Dutch influencers~(a). Word clouds of unique topics in the real dataset~(b), common topics~(c), and new topics in the synthetic data~(d) on Dutch influencers.}
    \label{fig:IDC_plt_aw}
\end{figure}

\begin{figure}[htb]
    \centering
    \includegraphics[width=0.75\linewidth]{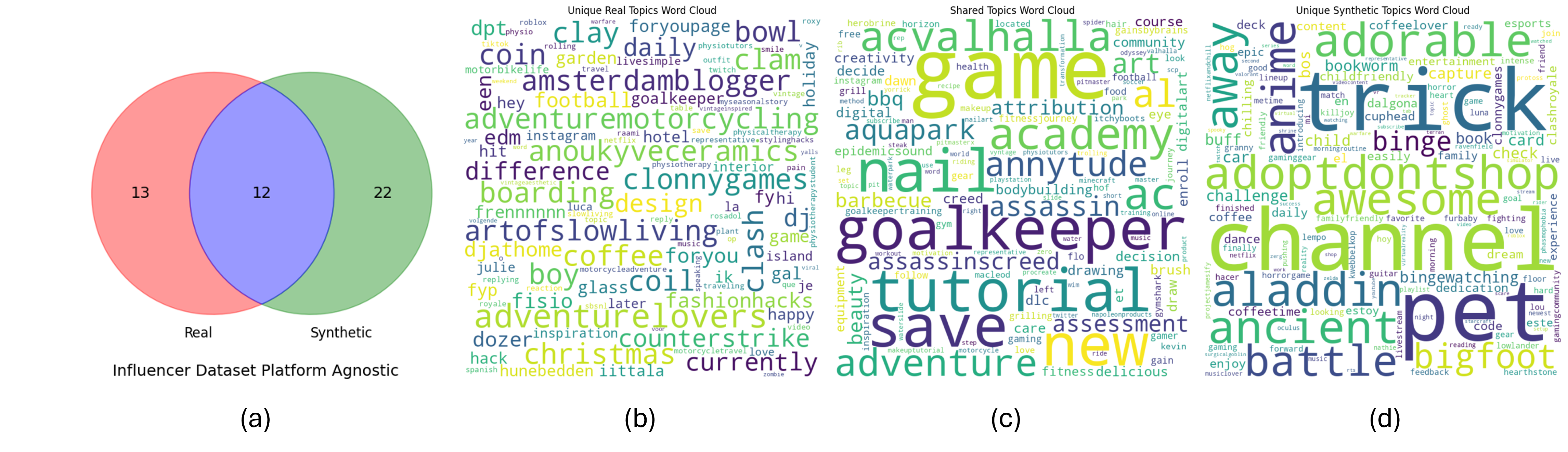}
    \caption{Topic overlap between real and synthetic (platform agnostic) data on the Dutch influencers~(a). Word clouds of unique topics in the real dataset~(b), common topics~(c), and new topics in the synthetic data~(d) on Dutch influencers.}
    \label{fig:IDC_plt_ag}
\end{figure}

\begin{figure}[htb]
    \centering
    \includegraphics[width=0.75\linewidth]{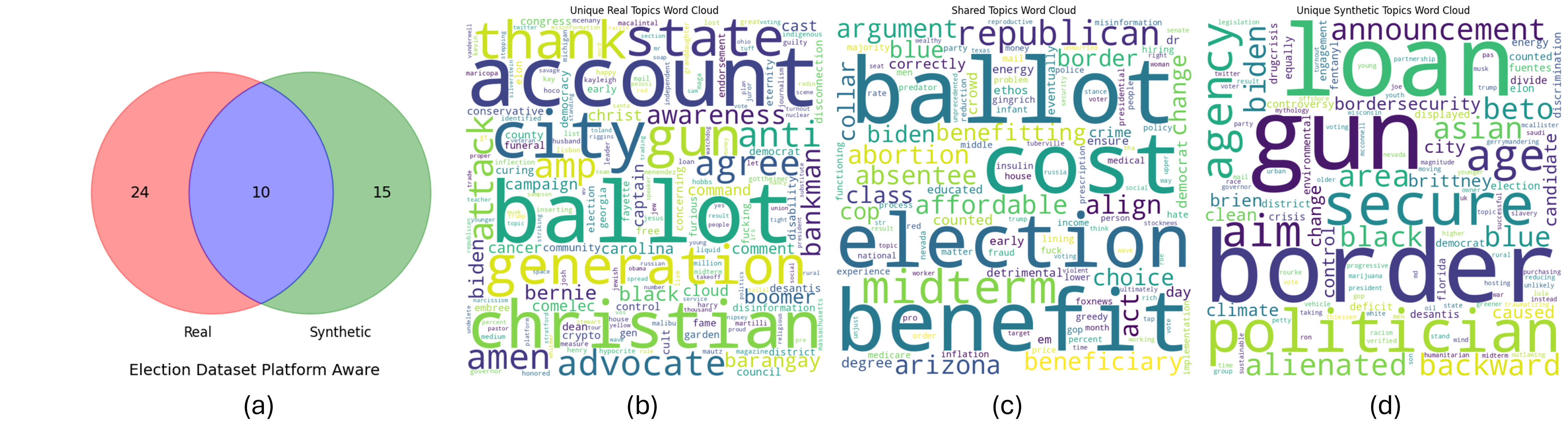}
    \caption{Topic overlap between real and synthetic (platform aware) data on the US elections~(a). Word clouds of unique topics in the real dataset~(b), common topics~(c), and new topics in the synthetic data~(d) on US elections.}
    \label{fig:EDC_plt_aw}
\end{figure}

Figures~\ref{fig:EDC_plt_ag},~\ref{fig:EDC_plt_aw},~\ref{fig:IDC_plt_ag}, and ~\ref{fig:IDC_plt_aw} present topic overlap between real and synthetic (both prompting) datasets along with word clouds of topic unique and common across datasets.
There are fewer topics in generated content in elections. 
In contrast, for the influencers data, generated content has more topics.
The word clouds are produced from the text generated by GPT 4 as descriptions of the top 10 words that describe a topic extracted with Bert. 
We observe that discussions in the synthetic data remain on the same general subjects as in the real datasets: for example, the synthetic US election datasets include state names (e.g., Arizona, Florida), typical election terminology (candidate, vote, ballot, absentee), politicians active in this context (Stefanik, Desantis, Feterman, Biden), and timely topics (border security, the opioid crisis, the release of Brittney Griner). 
Different topics and keywords have become more popular in the synthetic dataset, thus appearing in the word cloud: for example, ``Arabia'' in the word cloud representing topics unique to the synthetic data in Figure~\ref{fig:EDC_plt_ag} appears in the original dataset (in the context of oil prices) but it is emphasized significantly in the generated data.

In comparison, we observe more overlapping topics between platforms in the synthetic dataset of influencers.
Physiotherapy appears in synthetic data generated from both prompts on Instagram and YouTube. 
Amusement and games-related topics appear across all datasets in TikTok and YouTube.
In general, both synthetic influencers datasets stay on subjects specific to influencers' monetization practices, such as skin care, diets, life style (e.g., baking, chilling). 



\subsection{Embedding Similarity}
Most downstream tasks are solved by converting text into embedding vectors. Therefore, assessing the similarity of generated content with real content in embedding space becomes important.
To evaluate the similarity between real and synthetic social media posts, we used OpenAI's \textit{text-embedding-3-large} model~\cite{neelakantan2022text} to convert all text into embedding vectors. We then calculated pairwise cosine similarity for all possible pairs drawn from real and synthetic posts. To measure similarity, we selected the top 1000 highest similarity scores from each set and computed their average. This average represents the central tendency of the most similar pairs, providing a focused measure of similarity that highlights the best matches between the datasets. Additionally, we computed the average of all similarity scores across each dataset to establish a baseline, giving an overall sense of similarity regardless of the top matches. 
Table \ref{tab:res_embedding} displays the similarity matrix for platform agnostic and platform aware prompts with default settings in the election and the influencer dataset. 
In our experiment, platform-agnostic prompts yielded content with slightly higher similarity scores than the platform-aware prompt. Overall, both prompts generate content similar to the real dataset. Among the platforms, Facebook and YouTube exhibited the highest similarity for both prompting strategies, and Reddit and TikTok exhibited the lowest similarity. Our baseline showed that similarity in other cases is low at around 0.24.
\begin{table}[H]
\begin{tabular}{@{}l>{\centering\arraybackslash}p{1cm}>{\centering\arraybackslash}p{1cm}cc}
\toprule
\multirow{2}{*}{Platform} & \multicolumn{2}{c}{Platform Agnostic} & \multicolumn{2}{c}{Platform Aware}\\
\cmidrule(lr){2-3} \cmidrule(lr){4-5}
& \multicolumn{1}{c}{Top 1000} & Average& Top 1000&Average\\
\hline
Facebook                  & 0.824&0.234& 0.815&0.235\\
Twitter                   & 0.802&0.229& 0.804&0.233\\
Reddit                    &  0.762& 0.263& 0.754&0.261\\
 Instagran& 0.705& 0.201& 0.695&0.205\\
 TikTok& 0.693& 0.198& 0.682&0.210\\
 YouTube& 0.830& 0.273& 0.875&0.291\\
 \bottomrule
 \end{tabular}
\caption{Similarity matrix computed using cosine distance for Platform Agnostic and Platform Aware Prompt and T= 1, P= 1 parameters for Election and Influencer dataset.}
\label{tab:res_embedding}
\end{table}

\begin{figure}[!h]
\centering
\begin{subfigure}{0.45\textwidth}
\includegraphics[width=1\linewidth]{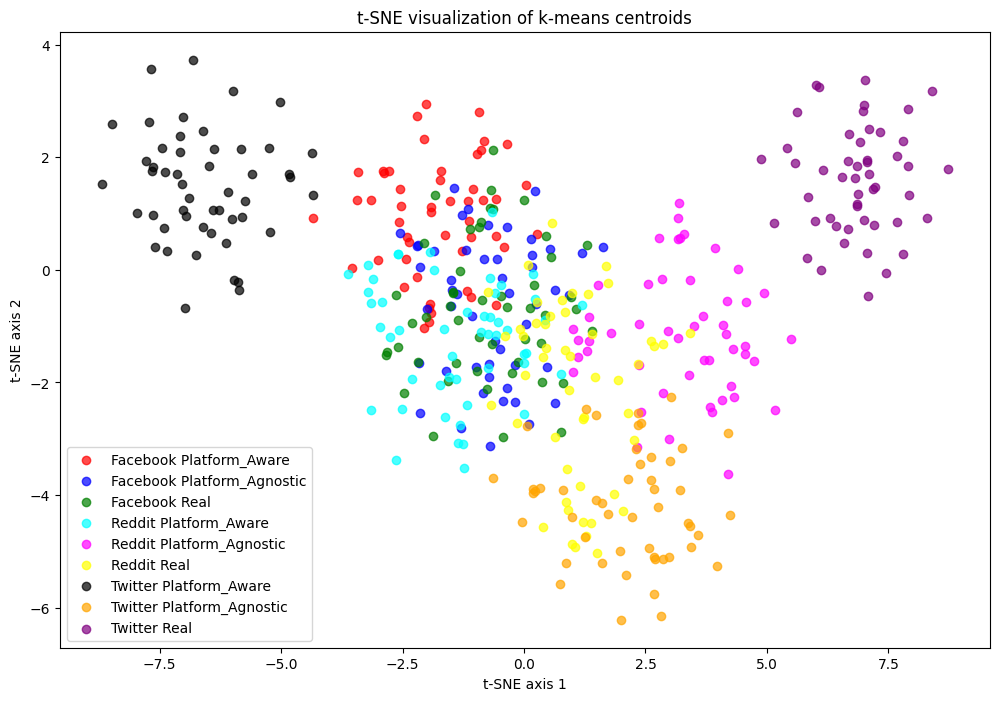}
    \caption{Elections dataset.}
\end{subfigure}
\begin{subfigure}{0.45\textwidth}
\includegraphics[width=1\linewidth]{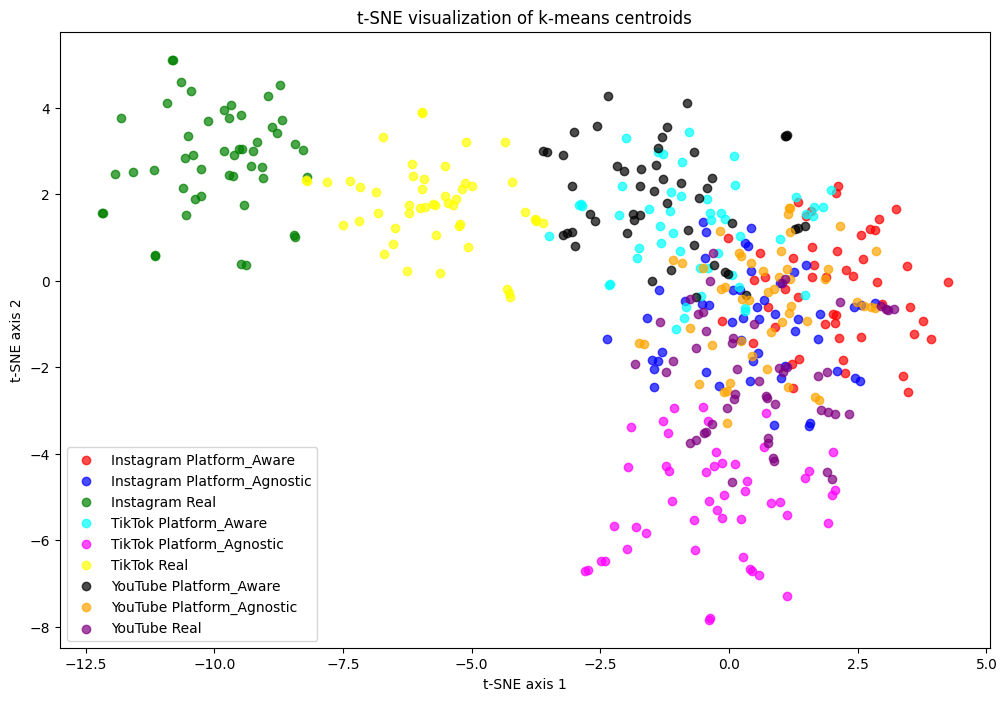}
\caption{Influencers dataset.}
\end{subfigure}
\caption{t-SNE plots of embedding vectors (1k data points are clustered into 50 clusters, and cluster centroids are plotted) of real, and synthetic (platform agnostic and platform aware).}
\label{fig:t-sne}
\end{figure}
To visualize embedding vectors, we created t-SNE plots using the embeddings of both real and synthetic datasets with the default settings (T=1, P=1). For each 1000 data points across all scenarios (platform aware, platform agnostic, and real), we apply K-means clustering (K=50) and represent cluster centroids as dots in Figure~\ref{fig:t-sne}. 
We experimented with different perplexity values and found consistent results, the plots are drawn on default perplexity value.
While t-SNE has limitations, it presents a good estimate for the preservation of local structure in data. 
In the election dataset, synthetic data shows high similarity with real data for Facebook.
In the case of Reddit and Twitter, synthetic datasets (platform aware and platform agnostic) show similarities along only one of the reduced dimensions. 
Real Reddit posts appear near to their synthetic counterparts, but for Twitter, they are quite far apart.
In the influencer dataset, we observe high similarity between real and synthetic video descriptions for YouTube.
For Instagram, both platform-aware and platform-agnostic datasets appear similar but quite separated from the real dataset. 
In the case of TikTok, only platform aware is aligned with real in one dimension, and others are quite far apart from each other. 
\section{Summary and Discussions}

This paper investigates the use of ChatGPT in generating social media datasets across various platforms and topics, focusing on two English-language datasets. The first dataset includes posts from Facebook, Reddit, and Twitter related to the US midterm 2022 elections, while the second comprises posts from TikTok, Instagram, and YouTube by a group of registered Dutch influencers.

We employed two prompting strategies: one where ChatGPT was prompted to generate posts similar to given examples without specifying the platform, and another where it was prompted to generate platform-specific posts. We also varied two parameters: temperature and probabilistic sampling. We assessed the quality of the generated datasets based on lexical features (hashtags, URLs, emojis, and user tags), sentiment, topics, and embedding similarity.

Our findings indicate that the synthetic datasets generally display high fidelity across various metrics. The synthetic posts notably preserve key social media lexical features such as emojis and hashtags and maintain semantic similarities with the original datasets, demonstrated through topic analysis and embedding similarity. However, neither prompting strategy consistently outperformed the other, and adjustments in parameters temperature and probabilistic sampling within our set limits did not significantly impact our evaluations.

Notable differences were observed in the utility of synthetic data for certain downstream tasks. For instance, ChatGPT-generated datasets underutilize tags and URLs compared to real datasets, with synthetic tags often not aligning with practices observed in human-generated content. Additionally, synthetic posts exhibited a disproportionately positive sentiment, particularly noticeable on Twitter, suggesting that other LLMs or alternative prompting strategies might better suit tasks requiring a broader range of sentiments.

The study has limitations that will be explored in future research. First, our results are based solely on the output of GPT-3.5-turbo as of early 2024, setting a benchmark for future studies that might replicate our approach with different LLMs. Second, as our datasets cover non-overlapping sets of platforms, it remains unclear whether the observed results are platform-specific or data-specific. Future studies will aim to explore various topics across the same platforms. Third, our focus was limited to English due to dataset availability and our linguistic proficiency, but future efforts will aim to include more diverse languages. Lastly, we did not address privacy and legal concerns related to releasing synthetic datasets.

This research demonstrates that generating synthetic social media datasets spanning multiple platforms is feasible with current technology. Improved prompting techniques and post-processing could enhance dataset fidelity. Realistic multi-platform synthetic datasets can promote reproducibility in social media research and offer resilience against restrictive platform policies, facilitating access to data for researchers globally.

\bibliographystyle{ACM-Reference-Format}
\bibliography{references}

\newpage

\section{Appendix}






\begin{table}[!htb]
\begin{center}
\caption{Comparison of topics (representative names of topics corresponding to topic words are generated through GPT4 prompting) in elections and influencers datasets. Fb: Facebook, Rd: Reddit, Tw: Twitter, In: Instagram, Tk: TikTok, and Yt: YouTube.}
\label{tab:inf_ele_topics}
\begin{tabular}{cccp{4cm}>{\raggedright\arraybackslash}p{4cm}>{\raggedright\arraybackslash}p{4cm}}
\toprule
\multicolumn{6}{c}{\textbf{Elections Dataset}} \\
\hline
 \textbf{Fb}  & \textbf{Rd} & \textbf{Tw} & \textbf{Real} & \textbf{Platform Agnostic}&\textbf{Platform Aware}\\
 \hline
 \checkmark & \checkmark & \checkmark & "Voter Engagement". & "Voter Engagement". &"Election Process and Voting Methods".\\
 \checkmark & \checkmark & & "Mail-in Voting and Absentee" and "Women Reproduction Rights" & "Mail-in Voting", "Legislative Control", "Social Media Misinformation" and
"Women Reproduction Rights". &"Women Reproduction Rights".\\
\midrule
\multicolumn{6}{c}{\textbf{Influencers Dataset}} \\
\hline
 \textbf{In}  & \textbf{Tk} & \textbf{Yt} & \textbf{Real} & \textbf{Platform Agnostic}&\textbf{Platform Aware}\\
 \hline
 \checkmark & \checkmark & \checkmark & "Holistic Health and Wellness Practice". & "Beauty and Style", "Culinary arts and Barbecue", "Health and Fitness" and "Live Stream Gaming". &"Fashion and Beauty Trends","Barbecue and Grilling Recipes" and "Fitness Routine and Motivation".\\
 \checkmark & \checkmark & & "Barbecue and Grilling Culinary Techniques" & "Streaming Series and Binge-Watching"  &"Hair Care and Styling", "Music and Live performances" and "Coffee Culture".\\
 \checkmark & & \checkmark & "Beauty
and Style" & "Physiotherapy education and Methods" and "Dig-
ital Art and Design"  &"Physiotherapy Education and Resources".\\
& \checkmark & \checkmark & "Amusement and Water
park Attractions" & "Video Games and
Consoles" &"Amusement and Theme Parks" and "Video Gaming and consoles".\\
 \bottomrule
\end{tabular}
\end{center}
\end{table}

\end{document}